# Interplay of Ferromagnetic and Antiferromagnetic Interactions in Epitaxial Co$_3$ZnN

Sita Dugu[1], Sharad Mahatara[1], Corlyn E Regier[2], James R Neilson[2], Stephan Lany[1], Rebecca W. Smaha[1], Sage R Bauers[1*]

## Affiliations

[1]Materials Sciences Center, National Renewable Energy Laboratory, Golden, Colorado 80401, United States

[2]Department of Chemistry, Colorado State University, Fort Collins, Colorado 80523-1872, United States

*Sage.Bauers@nrel.gov

## Abstract

Antiperovskite nitrides with the general formula $M_3A$N have attracted significant attention due to their tunable electronic and magnetic properties. Among them are many cobalt-based compounds predicted to exhibit high thermodynamic stability and intriguing magnetic behavior. Here, we report the synthesis and magnetic characterization of epitaxial Co$_3$ZnN thin films grown by radio frequency sputtering on SrTiO$_3$ (STO) and MgO substrates. X-ray diffraction confirms phase-pure (00$l$)-oriented films with cube-on-cube epitaxy on STO, with a $c$-lattice parameter of 3.752 Å. Magnetic measurements reveal clear hysteresis at 2 K with a coercive field of ~ 0.12 T and a small net moment of 0.11 $\mu_B$/f.u., suggesting either a canted antiferromagnetic (AFM) to ferrimagnetic (FiM) configuration. Temperature-dependent magnetization measurements show a transition near 25 K, with strong AFM interactions (Curie-Weiss $\Theta$ = –80.6 K) at high temperatures and short-range ferromagnetic correlations ($\Theta$ = +9.7 K) emerging near the transition. Complementary density functional theory (DFT) and Monte Carlo simulations indicate a ferromagnetic (FM) ground state, with the FM–AFM energy difference decreasing systematically with increasing supercell size, consistent with competition between FM and AFM/FiM interactions. These results highlight Co$_3$ZnN as a magnetically complex antiperovskite nitride with competing exchange interactions.

Antiperovskite (AP) nitrides with the general formula $M_3A$N represent a significant branch of all $M_3AX$ AP compounds and are known for their diverse and intriguing physical properties as well as their potential multifunctionalities. [1], [2] Within this family, several subclasses with 3$d$ $M$-site metals have been extensively studied. For instance, Ni-based nitride Ni$_3A$N ($A$ = Cu and Zn) show superconducting behavior [3], [4]; Fe-based systems like Fe$_3A$N ($A$ = Pd, Mn, Fe, Sn and Ni) show nearly invariant coefficients of thermal expansion (CTEs) [5], [6], [7], [8], [9], [10], [11]; Mn-based compounds, including Mn$_3A$N ($A$ = Zn, Ge, Ga, and Cu) display large negative CTEs [12], [13], magnetocaloric and barocaloric effects [14], [15], spin glass (SG) behavior [16], [17], and magnetostriction [18]; and Cr-based nitrides Cr$_3A$N ($A$ = Ga, Pd, Rh, and Pt) have been reported to exhibit superconductivity, SG behavior and zero-field cooled (ZFC) exchange bias effects [19], [20].

Despite several experimental studies on $M_3A$N compounds, magnetic investigations of Co-based AP nitrides remain limited due to challenges in synthesizing high-quality samples such as single crystals or epitaxial films. Yet high-throughput density functional theory (DFT) calculations by Singh et al. [21] indicate that Co-based cubic APs are thermodynamically stable, lying on the convex hull. Recent experiments support the stability of Co-based APs and provide insight into their magnetic behavior. For example, Liu *et al.* [22] reported spin-glass behavior in Co$_3$SnN attributed to competing antiferromagnetic (AFM) and ferromagnetic (FM) interactions arising from atomic disorder. Co$_3$FeN displays reversible magnetic domain structures and magnetization, and studies on bilayer thin films combining FM Co$_3$FeN with AFM Mn$_3$GaN demonstrate current-induced spin-transfer torque effects in the Mn$_3$GaN layer [23].

Our prior work demonstrated the growth of another theoretically predicted Co-based AP, Co$_3$PdN.[24] Motivated by the successful synthesis of Co$_3$PdN and its high-temperature ferromagnetism (Curie temperature of $T_C \approx 560$ K), here we explore another candidate material, Co$_3$ZnN, identified in high-throughput DFT as lying on the convex hull of stability.[21] While phase-pure Co$_3$ZnN ($a$ = 3.764 Å) was recently synthesized via ammonolysis for nitrogen storage applications [25], no studies have been reported thin-film fabrication or magnetic characterization.

Here, we report the epitaxial growth of Co$_3$ZnN films using RF co-sputtering and investigate their magnetic properties. We grew the Co$_3$ZnN films on strontium titanate (STO) and magnesium oxide (MgO) substrates by radio frequency (RF) co-sputtering in a mixture of Ar and N$_2$ gases. Uniform films with a final composition of Co$_{2.97}$Zn$_{1.01}$N$_{0.95}$O$_{0.07}$, determined using electron probe microanalysis (EPMA), were achieved by applying 70 watts to the Co and 40 watts to the Zn sources (both >99.9% pure).

X-ray diffraction (XRD) patterns of the films are shown in Figure 1a, where only (00$l$) reflections are observed. In nitride APs, the odd (00$l$) reflections arise primarily due to the difference in scattering factors between the corner (Zn) and face-center (Co) atoms. Thus, the (001) peak (shown in inset of Figure 1a) exhibits low intensity primarily because of the similar Z values of Co and Zn. Co$_3$ZnN films grown on MgO (Figure S1 in the Supplementary Material) also show exclusively (00$l$) reflections. This is consistent with the expected cubic $Pm\bar{3}m$ symmetry of the AP structure. We calculate the out-of-plane lattice parameter to be 3.752 Å and found no indication of tetragonal distortion. The lattice mismatch of the film is +3.56% on STO, and it increases to +10.92% on MgO, so we selected STO for further growth and characterization. Figure 1b shows an XRD pole figure collected from the (110) plane of Co$_3$ZnN. The discrete points of intensity overlap in $\chi/\varphi$ with the substrate's (110) poles, indicating cube-on-cube epitaxy. φ-scans collected from the film and STO substrate from [111] planes (Figure 1c) also show peaks separated by 90°, consistent with the expected fourfold symmetry. This confirms a cube-on-cube relationship between the Co$_3$ZnN film and the (001)-oriented STO substrate (Figure 1d), establishing the in-plane epitaxial relationship as Co$_3$ZnN(001)[100] ∥ STO(001)[100].

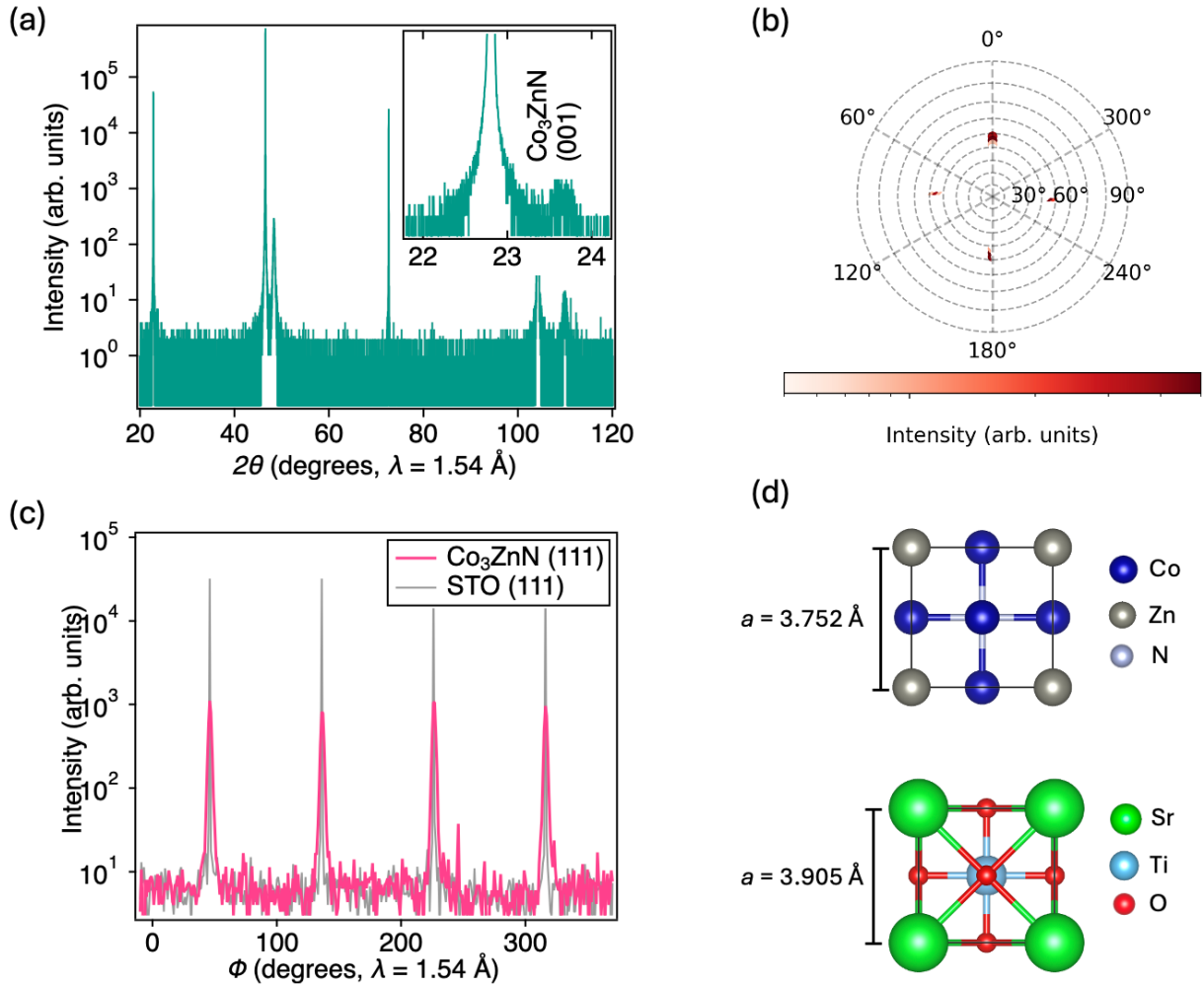

Figure 1: (a) presents XRD data collected in $\omega$-$2\theta$ geometry from Co$_3$ZnN thin films grown on (00$l$) STO substrate. (b) Pole figure of the Co$_3$ZnN (110) family of peaks. (c) $\varphi$-scans collected at the (111) Bragg condition for Co$_3$ZnN and STO. (d) Crystal structure schematics of antiperovskite Co$_3$ZnN and perovskite STO.

We characterized the magnetic properties of epitaxial $Co_3ZnN$ films on STO substrates using DC magnetization and SQUID magnetometry. Figure 2(a) shows magnetic hysteresis loops measured at various temperatures; the data are plotted after subtracting the diamagnetic contribution from the substrates (see Figure S2 in the Supplementary Material). A clear hysteresis is observed at 2 K, which disappears above 20 K. The film exhibits a coercive field ($H_C$) ~ 0.12 T, significantly higher than that of polycrystalline $Co_3PdN$ reported in our prior investigation [24]. However, the remanent magnetization at 2 K is approximately 20 emu/cm$^3$ (0.11 $\mu_B$/f.u.), which is notably smaller than that of $Co_3PdN$ (~550 emu/cm$^3$). The presence of a hysteresis loop with small net magnetization suggests that the system likely has either a canted AFM or a ferrimagnetic (FiM) ground state. The raw data of the sample including the substrate contribution are shown in Figure S2a in the Supplementary Material. Figure S2b displays the raw data of a bare STO substrate.

Temperature-dependent magnetization was measured under both zero-field-cooled (ZFC) and field-cooled (FC) conditions at applied fields of 0.05 T and 0.5 T, as shown in Figure 2b. To estimate the transition temperature, the first derivative of the FC curve collected at 0.05 T was calculated, revealing a transition around 25 K (Figure 2c). The sharp decrease in the ZFC curve below the transition suggests AFM ordering. A pronounced bifurcation between the ZFC and FC curves is observed at low temperatures under an applied field of 0.05 T with a sharp peak in the ZFC curve at ~20 K. This bifurcation is not present in the data collected at 0.5 T. This behavior may indicate some spin glass character of the ground state, which was previously reported in $Co_3SnN$, [22], [26] although it is also relatively common in complex antiferromagnets.

To further investigate the magnetic ordering, the FC data collected at 0.05 T were fitted to the Curie–Weiss law [27], as shown in Figure 2d. A fit in the 100–300 K range yields a Weiss temperature ($\Theta$) of –80.6(3) K, indicating strong AFM interactions in the paramagnetic regime. Overall, this is a sharp departure from the magnetic behavior observed for $Co_3PdN$ films, which exhibit strong ferromagnetism with a $T_C$ of 563.2(4) K.[24] However, antiferromagnetism or ferrimagnetism are quite common in the $M_3A$N AP family as a whole, with many Mn-based compounds displaying AFM or FiM ground states.[2]

Interestingly, Figure 2d shows a pronounced kink in the inverse moment data at approximately 60 K. When the temperature-dependent magnetization data are fitted below this kink yet above the transition temperature (25–50 K), the extracted $\Theta$ is +9.7(3) K. This small positive value suggests the emergence of some short-range FM correlations near the transition temperature, supporting the appearance of hysteresis below the transition. Altogether, these results imply that $Co_3ZnN$ exhibits either a canted AFM ground state or a FiM ground state in which there is coexistence of FM and AFM ordering likely driven by competing exchange interactions.

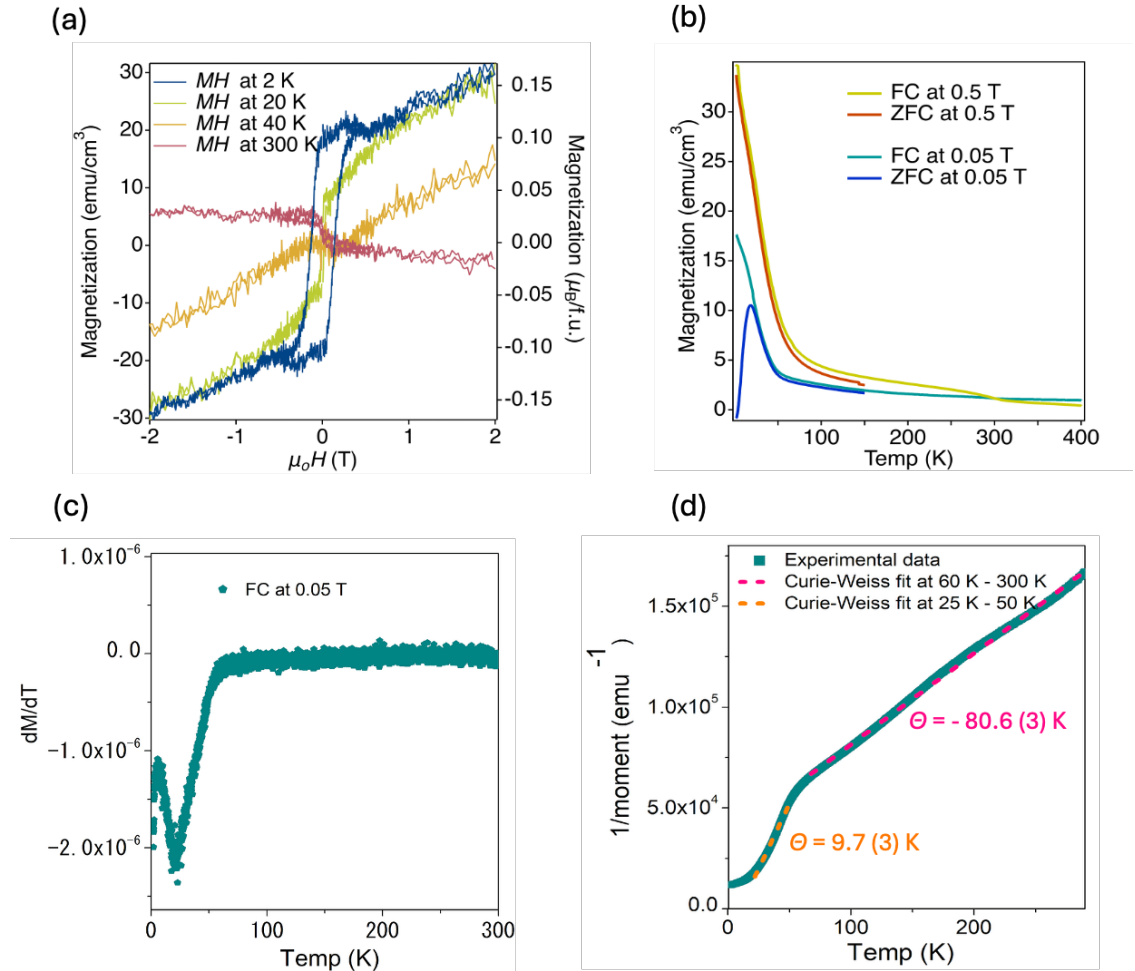

Figure 2(a) Magnetization versus field of Co$_3$ZnN film measured from 2 K – 300 K. (b) Magnetization versus temperature measured at 0.05 T and 0.5 T for zero-field cooled and field cooled condition. (c) 1st order derivative of magnetization w.r.t temperature calculated for field cooled data measured at 0.05 T. (d) Inverse moment as a function of temperature with Curie-Weiss fits in various temperature ranges for field cooled data measured at 0.05 T.

Similar competition between AFM and FM interactions has been reported in Co$_3$GeN [28] and Co$_3$SnN,[22] where it is ascribed to SG effects. In Co$_3$GeN, a canonical SG state arises from atomic disorder caused by Ge vacancies, with measured Co:Ge ratio of 3.0:0.82. Likewise, in Co$_3$SnN, SG features are linked either to Sn deficiency or the coexistence of Co-Sn ferromagnetic clusters and Co-N antiferromagnetic interactions, or a combination of both effects. For Co$_3$ZnN, however, our EPMA measurements indicate a global composition very close to stoichiometric (Co$_{2.97}$Zn$_{1.01}$N$_{0.95}$O$_{0.07}$), suggesting that a large-scale deviation from the expected 3:1:1 stoichiometry is unlikely. Moreover, the presence of the (001) peak in the PXRD data (Figure 1a) implies that the Co and Zn atoms are ordered on their specific sites within the cubic AP structure.

To determine the magnetic ground state, we performed DFT calculations on atomically ordered cubic Co$_3$ZnN using a supercell approach. We compared collinear FM and AFM configurations using a 10-atom face-centered cubic (fcc) supercell. In this 10-atom cell, the FM state is lower by

30.8 meV/Co than the AFM state considering all 10 collinear AFM possibilities; the Co local moment is ~1 $\mu_B$ and the total magnetization is ~3 $\mu_B$/f.u. We note that other GGA studies also report an FM ground state.[29], [30] Further, we extended the study to 20-, 40-, and 60-atom supercells and used first-principles Monte Carlo sampling on DFT (based on GGA) energies to search for ground state magnetic configurations (see Methods in the Supplementary Material). In the 20-atom cell, FM is favored by 26.7 meV per Co, while in the 40-atom cell the energy difference decreases to 12.0 meV per Co; in both larger supercells the AFM solutions converge to zero net magnetization. A larger 60-atom supercell was further modeled, showing a reduced energy difference of 8.9 meV per Co, with the initial AFM configuration relaxing to a FiM state. The systematic reduction in the FM–AFM/FiM energy gap with increasing cell size indicates consistent FM ordering but increasing AFM/FiM competition, consistent with frustrated long-range exchange interactions, as shown in Figure 3a.

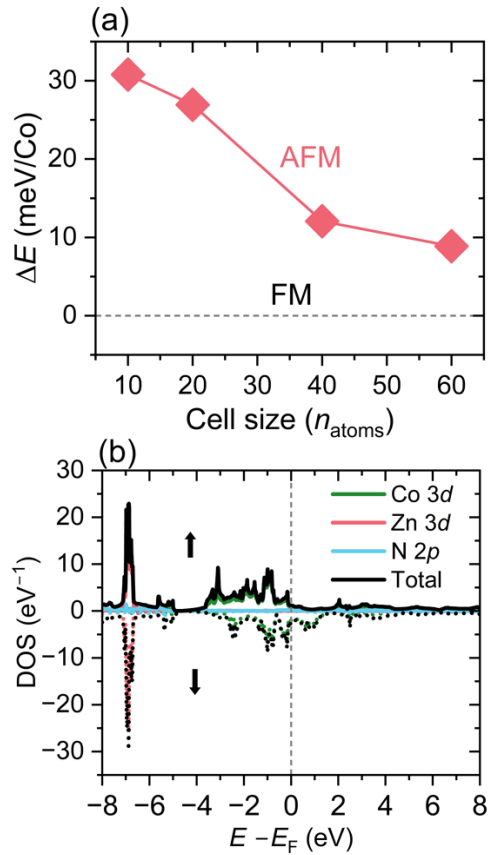

Figure 3(a) The collinear energy differences, calculated relative to the FM ground state, show that the competition between FM and AFM states increases with supercell size and (b) the DOS of $Co_3ZnN$ for FM ground state.

To explore possible non-collinear magnetic states, we performed GGA non-collinear + spin orbit coupling (SOC) calculations for several candidate structures (AFM-1 $\Gamma_{5g}$,[31] AFM-2 $\Gamma_{4g}$,[32] AFM-3, AFM-4, and a FiM 1:2 up:down configuration), as summarized in Table 1. All six non-collinear AFM or FiM configurations are higher in energy than the FM state; the lowest, AFM-4, lies ~20.8 meV per Co above the FM configuration (compared to 12.0 meV per Co for the collinear 40-atom cell). Thus, within GGA, we can rule out AFM or FiM ground states while recognizing FM–AFM competition, consistent with the collinear results in Fig. 3a and the experimental behavior in Fig. 2. The low transition temperature ($T_N$ = 25 K), small net moment (0.11 $\mu_B$/f.u.), and weak FM correlations just above the transition ($\Theta$ = +9.7(3) K) in Co$_3$ZnN support an AFM ground state with competing FM interactions. While strain from growth on STO could induce tetragonal distortion, our films are relaxed; a low-temperature structural transition, however, could modify the energetics, potentially reconciling the observed canted AFM or FiM behavior with the DFT-predicted FM ground state. To further assess possible structural transitions, we performed structure prediction calculations at the GGA level.[33] The FM cubic antiperovskite structure is confirmed as the ground state in both 10- and 20-atom searches.

Table1: Relative energies of non-collinear magnetic configurations of Co$_3$ZnN from GGA+SOC calculations.

| Magnetic structure | Super-cell size (atoms) | $\Delta E$ (meV/Co) |
|---|---|---|
| FM | 5 | 0 |
| AFM-1, $\Gamma_{5g}$ | 5 | 27.5 |
| AFM-2, $\Gamma_{4g}$ | 5 | 35.8 |
| AFM-3 | 10 | 30.7 |
| FiM | 10 | 30.8 |
| AFM-4 | 40 | 20.8 |

Overall, the structure search, collinear DFT-based Monte Carlo simulations, and non-collinear calculations consistently indicate a FM ground state. However, we recognize that this is not consistent with the experimental results that indicate canted AFM or FiM order at 2 K. We note that the energy competition observed in larger supercells, as shown in Fig. 3a, suggests the possible emergence of more complex AFM or FiM orders at longer length scales. In addition, non-collinear canting, which may not be fully captured in our calculations, could also lead to a magnetic ground state distinct from the ideal FM configuration. Either—or both—of these effects may be at play in this system, and it is also possible that the cubic structure used for these calculations may not be an accurate representation of the structure of Co$_3$ZnN near or below its magnetic transition as the SrTiO$_3$ substrate, which undergoes a tetragonal phase transition at 105 K, may strain the film.

As shown in Figure 3b, the calculated density of states (DOS) confirms the metallic nature of Co$_3$ZnN. The asymmetric distribution of the spin-resolved DOS near the Fermi energy ($E_F$), with a larger contribution from the down-spin channel, resembles the behavior reported for Co$_3$PdN.[24] The Co 3$d$ electrons predominantly contribute to the conduction band and exhibit a high DOS near $E_F$, underscoring their primary role in the electronic and magnetic properties of Co$_3$ZnN. The Co atoms are primarily responsible for the observed magnetism. In contrast, the Zn $d$ states lie far below $E_F$, suggesting weak electronic hybridization between Co $d$, N $p$ and Zn $d$ states, and consequently, weaker FM interactions compared to Co$_3$PdN.[24]

In conclusion, we synthesized and then successfully studied the structural and magnetic properties of epitaxial Co$_3$ZnN films. X-ray diffraction confirmed phase-pure (00$l$)-oriented films with cube-on-cube epitaxy and a lattice parameter of 3.752 Å. Magnetic measurements revealed hysteresis at 2 K with a coercive field of ~ 0.12 T and a small net moment, indicating a canted AFM or FiM ground state. Temperature-dependent magnetization identified a transition near 25 K and strong AFM interactions with short-range FM correlations. DFT and Monte Carlo simulations predicted a FM ground state, but the narrowing FM−AFM energy difference with increasing cell size suggests competing magnetic interactions. Overall, Co$_3$ZnN exhibits complex magnetism arising from the interplay of FM and AFM exchange couplings. Idealized DFT favors FM, but real thin-film materials with subtle disorder, distortions, or long-range competing exchange relax into a canted AFM or FiM state.

## SUPPLEMENTARY MATERIAL

See the supplementary material for detailed methods and additional XRD and magnetization data.

## AUTHOR DECLARATIONS

### Conflict of Interest

There are no conflicts to declare.

### Author Contributions

**Sita Dugu:** Conceptualization (equal); Formal analysis (equal); Methodology (equal); Investigation (equal); Software (equal); Writing – original draft (equal). **Sharad Mahatara:** Methodology (equal); Software (equal); Writing – original computational draft (equal). **Corlyn E. Regier:** Methodology-magnetic measurement (equal). **James R. Neilson:** Visualization (equal); Writing – review & editing (equal). **Rebecca Smaha:** Visualization (equal); Writing – review & editing (equal). **Stephan Lany:** Supervision (equal); Visualization (equal); Writing – review & editing (equal). **Bauers:** Conceptualization (equal); Supervision (equal); Visualization (equal); Writing – review & editing (equal).

## DATA AVAILABILITY

The data that support the findings of this study are available from the corresponding authors upon reasonable request. Experimental synthesis data are also publicly available in the National Renewable Energy Laboratory (NREL) high-throughput experimental materials database.[34]

## ACKNOWLEDGEMENTS

This work was authored by the National Laboratory of the Rockies (NLR), operated by Alliance for Sustainable Energy, LLC, for the U.S. Department of Energy (DOE) under Contract No. DE-AC36-08GO28308. Funding provided by the U.S. Department of Energy, Office of Science, Basic Energy Sciences, Division of Materials Science, through the Office of Science Funding Opportunity Announcement (FOA) Number DE- FOA-0002676: Chemical and Materials Sciences to Advance Clean-Energy Technologies and Transform Manufacturing. The research used High-Performance Computing (HPC) resources of the National Energy Research Scientific Computing Center (NERSC), a DOE-SC user facility located at Lawrence Berkeley National Laboratory, operated under Contract No. DE-AC02-05CH11231. This research also used HPC resources at NLR, sponsored by DOE, Office of Critical Minerals and Energy Innovation. The views expressed in the article do not necessarily represent the views of the DOE or the U.S. Government. The authors thank the Analytical Resources Core at Colorado State University for instrument access and training (RRID: SCR_021758).

## REFERENCES

[1] Y. Wang et al., "Antiperovskites with Exceptional Functionalities," *Adv. Mater.*, vol. 32, no. 7, p. 1905007, 2020, doi: 10.1002/adma.201905007.

[2] R. Niewa, "Metal-Rich Ternary Perovskite Nitrides," *Eur. J. Inorg. Chem.*, vol. 2019, no. 32, pp. 3647–3660, 2019, doi: 10.1002/ejic.201900756.

[3] B. He et al., "CuNNi3: a new nitride superconductor with antiperovskite structure," *Supercond. Sci. Technol.*, vol. 26, no. 12, p. 125015, Nov. 2013, doi: 10.1088/0953-2048/26/12/125015.

[4] M. Uehara, A. Uehara, K. Kozawa, T. Yamazaki, and Y. Kimishima, "New antiperovskite superconductor ZnNNi3, and related compounds CdNNi3 and InNNi3," *Phys. C Supercond. Its Appl.*, vol. 470, pp. S688–S690, Dec. 2010, doi: 10.1016/j.physc.2009.11.131.

[5] S. Matar, P. Mohn, G. Demazeau, and B. Siberchicot, "The calculated electronic and magnetic structures of Fe4N and Mn 4N," *J. Phys.*, vol. 49, no. 10, pp. 1761–1768, Oct. 1988, doi: 10.1051/jphys:0198800490100176100.

[6] "Magnetic particles derived from iron nitride." Accessed: July 16, 2025. [Online]. Available: https://xplorestaging.ieee.org/document/50490?denied=

[7] C. A. Kuhnen and A. V. dos Santos, "Magnetic and electronic structure of the nitrides PdFe3N and MnFe3N," *J. Magn. Magn. Mater.*, vol. 130, no. 1, pp. 353–362, Feb. 1994, doi: 10.1016/0304-8853(94)90694-7.

[8] P. Mohn, K. Schwarz, S. Matar, and G. Demazeau, "Calculated electronic and magnetic structure of the nitrides $\mathrm{NiFe}_{3}$N and $\mathrm{PdFe}_{3}$N," *Phys. Rev. B*, vol. 45, no. 8, pp. 4000–4007, Feb. 1992, doi: 10.1103/PhysRevB.45.4000.

[9] C. A. Kuhnen and A. V. dos Santos, "Calculated electronic structure of SnFe3N," *Solid State Commun.*, vol. 85, no. 3, pp. 273–279, Jan. 1993, doi: 10.1016/0038-1098(93)90452-S.

[10] S. Suzuki, H. Sakumoto, J. Minegishi, and Y. Omote, "Coercivity and unit particle size of metal pigment," *IEEE Trans. Magn.*, vol. 17, no. 6, pp. 3017–3019, Nov. 1981, doi: 10.1109/TMAG.1981.1061756.

[11] C. A. Kuhnen, R. S. de Figueiredo, V. Drago, and E. Z. da Silva, "Mössbauer studies and electronic structure of γ′-Fe4N," *J. Magn. Magn. Mater.*, vol. 111, no. 1, pp. 95–104, June 1992, doi: 10.1016/0304-8853(92)91062-X.

[12] K. Takenaka and H. Takagi, "Giant negative thermal expansion in Ge-doped anti-perovskite manganese nitrides," *Appl. Phys. Lett.*, vol. 87, no. 26, p. 261902, Dec. 2005, doi: 10.1063/1.2147726.

[13] D. Fruchart and E. F. Bertaut, "Magnetic Studies of the Metallic Perovskite-Type Compounds of Manganese," *J. Phys. Soc. Jpn.*, vol. 44, no. 3, pp. 781–791, Mar. 1978, doi: 10.1143/JPSJ.44.781.

[14] J. Yan et al., "Phase transitions and magnetocaloric effect in Mn3Cu0.89N0.96," *Acta Mater.*, vol. 74, pp. 58–65, Aug. 2014, doi: 10.1016/j.actamat.2014.04.005.

[15] D. Matsunami, A. Fujita, K. Takenaka, and M. Kano, "Giant barocaloric effect enhanced by the frustration of the antiferromagnetic phase in Mn3GaN," *Nat. Mater.*, vol. 14, no. 1, pp. 73–78, Jan. 2015, doi: 10.1038/nmat4117.

[16] B. Song *et al.*, "Observation of spin-glass behavior in antiperovskite Mn3GaN," *Appl. Phys. Lett.*, vol. 92, no. 19, p. 192511, May 2008, doi: 10.1063/1.2931058.
[17] X. H. Zhang *et al.*, "Observation of spin-glass behavior in antiperovskite compound Mn3Cu0.7Ga0.3N," *Appl. Phys. Lett.*, vol. 103, no. 2, p. 022405, July 2013, doi: 10.1063/1.4813412.
[18] K. Asano, K. Koyama, and K. Takenaka, "Magnetostriction in Mn3CuN," *Appl. Phys. Lett.*, vol. 92, no. 16, p. 161909, Apr. 2008, doi: 10.1063/1.2917472.
[19] B. Wiendlocha, J. Tobola, S. Kaprzyk, and D. Fruchart, "Electronic structure, superconductivity and magnetism study of Cr3GaN and Cr3RhN," *J. Alloys Compd.*, vol. 442, no. 1, pp. 289–291, Sept. 2007, doi: 10.1016/j.jallcom.2006.08.365.
[20] S. Lin *et al.*, "Spin-glass behavior and zero-field-cooled exchange bias in a Cr-based antiperovskite compound PdNCr3," *J. Mater. Chem. C*, vol. 3, no. 22, pp. 5683–5696, May 2015, doi: 10.1039/C5TC00423C.
[21] H. K. Singh, Z. Zhang, I. Opahle, D. Ohmer, Y. Yao, and H. Zhang, "High-Throughput Screening of Magnetic Antiperovskites," *Chem. Mater.*, vol. 30, no. 20, pp. 6983–6991, Oct. 2018, doi: 10.1021/acs.chemmater.8b01618.
[22] C. Liu *et al.*, "Spin-glass behavior in Co-based antiperovskite compound SnNCo3," *Appl. Phys. Lett.*, vol. 116, no. 5, p. 052401, Feb. 2020, doi: 10.1063/1.5140434.
[23] H. Sakakibara, H. Ando, Y. Kuroki, S. Kawai, K. Ueda, and H. Asano, "Magnetic properties and anisotropic magnetoresistance of antiperovskite nitride Mn3GaN/Co3FeN exchange-coupled bilayers," *J. Appl. Phys.*, vol. 117, no. 17, p. 17D725, May 2015, doi: 10.1063/1.4917501.
[24] S. Dugu *et al.*, "Synthesis, Stability, and Magnetic Properties of Antiperovskite Co3PdN," *Chem. Mater.*, vol. 37, no. 5, pp. 1906–1913, Mar. 2025, doi: 10.1021/acs.chemmater.4c03147.
[25] Y. Goto, A. Daisley, and J. S. J. Hargreaves, "Towards anti-perovskite nitrides as potential nitrogen storage materials for chemical looping ammonia production: Reduction of Co3ZnN, Ni3ZnN, Co3InN and Ni3InN under hydrogen," *Catal. Today*, vol. 364, pp. 196–201, Mar. 2021, doi: 10.1016/j.cattod.2020.03.022.
[26] "Glassy ferromagnetism in -type | Phys. Rev. B." Accessed: Oct. 15, 2025. [Online]. Available: https://journals.aps.org/prb/abstract/10.1103/PhysRevB.73.205105
[27] L. Zu *et al.*, "A first-order antiferromagnetic-paramagnetic transition induced by structural transition in GeNCr3," *Appl. Phys. Lett.*, vol. 108, no. 3, p. 031906, Jan. 2016, doi: 10.1063/1.4940365.
[28] L. Zu *et al.*, "Observation of the Spin-Glass Behavior in Co-Based Antiperovskite Nitride GeNCo3," *Inorg. Chem.*, vol. 55, no. 18, pp. 9346–9351, Sept. 2016, doi: 10.1021/acs.inorgchem.6b01462.
[29] H. Hu, X. Yan, X. Wang, C. Yin, J. P. Attfield, and M. Yang, "Magnetic Properties and Electrocatalytic Oxygen Evolution Performance of a Medium-Entropy Metal Nitride," *Chem. Mater.*, vol. 36, no. 23, pp. 11432–11439, Dec. 2024, doi: 10.1021/acs.chemmater.4c02059.
[30] S. Liu *et al.*, "Efficient photocatalytic hydrogen evolution over carbon supported antiperovskite cobalt zinc nitride," *Chem. Eng. J.*, vol. 408, p. 127307, Mar. 2021, doi: 10.1016/j.cej.2020.127307.
[31] G. Gurung, M. Elekhtiar, Q.-Q. Luo, D.-F. Shao, and E. Y. Tsymbal, "Nearly perfect spin polarization of noncollinear antiferromagnets," *Nat. Commun.*, vol. 15, no. 1, p. 10242, Nov. 2024, doi: 10.1038/s41467-024-54526-1.
[32] Z. Miao, X. Guo, G. Yang, N. Wang, and J. Li, "The spin-glass-like behavior in antiperovskite Mn3AgN," *J. Appl. Phys.*, vol. 136, no. 5, p. 055104, Aug. 2024, doi: 10.1063/5.0217465.
[33] A. Sharan and S. Lany, "Computational discovery of stable and metastable ternary oxynitrides," *J. Chem. Phys.*, vol. 154, no. 23, p. 234706, June 2021, doi: 10.1063/5.0050356.
[34] A. Zakutayev *et al.*, "An open experimental database for exploring inorganic materials," *Sci. Data 2018 51*, vol. 5, no. 1, pp. 1–12, Apr. 2018, doi: 10.1038/sdata.2018.53.
[35] K. R. Talley *et al.*, "Synthesis of Lanthanum Tungsten Oxynitride Perovskite Thin Films," *Adv. Electron. Mater.*, vol. 5, no. 7, p. 1900214, 2019, doi: 10.1002/aelm.201900214.
[36] "Search | HTEM." Accessed: Nov. 11, 2025. [Online]. Available: https://htem.nrel.gov/

[37] S. K. Wallace, J. M. Frost, and A. Walsh, "Atomistic insights into the order–disorder transition in Cu2ZnSnS4 solar cells from Monte Carlo simulations," *J. Mater. Chem. A*, vol. 7, no. 1, pp. 312–321, Dec. 2018, doi: 10.1039/C8TA04812F.

[38] K. Binder, *Monte Carlo simulation in statistical physics: an introduction*, 3rd ed. in Springer series in solid-state sciences, 80. Berlin ; Springer, 1997.

[39] J. P. Perdew, K. Burke, and M. Ernzerhof, "Generalized Gradient Approximation Made Simple," *Phys. Rev. Lett.*, vol. 77, no. 18, pp. 3865–3868, Oct. 1996, doi: 10.1103/PhysRevLett.77.3865.

[40] P. E. Blöchl, "Projector augmented-wave method," *Phys. Rev. B*, vol. 50, no. 24, pp. 17953–17979, Dec. 1994, doi: 10.1103/PhysRevB.50.17953.

[41] G. Kresse and J. Furthmüller, "Efficient iterative schemes for ab initio total-energy calculations using a plane-wave basis set," *Phys. Rev. B*, vol. 54, no. 16, pp. 11169–11186, Oct. 1996, doi: 10.1103/PhysRevB.54.11169.

[42] H. Peng, D. O. Scanlon, V. Stevanovic, J. Vidal, G. W. Watson, and S. Lany, "Convergence of density and hybrid functional defect calculations for compound semiconductors," *Phys. Rev. B*, vol. 88, no. 11, p. 115201, Sept. 2013, doi: 10.1103/PhysRevB.88.115201.

# Supplementary Material

# Interplay of Ferromagnetic and Antiferromagnetic Interactions in Epitaxial Co$_3$ZnN

Sita Dugu[1], Sharad Mahatara[1], Corlyn E Regier[2], James R Neilson[2], Rebecca Smaha[1], Stephan Lany[1], Sage R Bauers[1]

[1]Materials Sciences Center, National Renewable Energy Laboratory, Golden, Colorado 80401, United States

[2]Department of Chemistry, Colorado State University, Fort Collins, Colorado 80523-1872, United States

# Methods

## Material synthesis

Co and Zn targets (99.99%, Kurt J. Lesker) with 50.8 mm diameter were co-sputtered to achieve stoichiometric Co$_3$ZnN antiperovskite thin films on strontium titanate (STO) (*00l*) and magnesium oxide (MgO) (*00l*) crystal substrate (both from MTI Corp.). The power densities of Co (2.7 W/cm$^2$) and Zn (1.5 W/cm$^2$) were optimized through a series of growth on Si substrates. A reactive gas mixture of N$_2$ (6 sccm) and Ar (1 sccm) was introduced in such a way that the pressure in the chamber was maintained at 1.5 mTorr using a throttle valve. The substrate was rotated to achieve uniform films. Film growth was carried out at a substrate temperature of 250 °C for 10 mins for subsequent structural analysis. X-ray diffraction (XRD) measurements were performed on a SmartLab 3kW X-ray diffractometer using monochromatic Cu-k$\alpha$ radiation, with the goniometer aligned to the substrate crystal. Experimental data used by this study have been analyzed using COMBIgor software [35] was used for data analysis and figure preparation on the experimental data used on this study and are publicly available in the National Renewable Energy Laboratory (NREL) high-throughput experimental materials database [36].

## Material Characterization

Magnetic properties were measured on a thicker film that was grown for 30 minutes. Hysteresis data were measured in a Quantum Design Physical Property Measurement System as DC magnetization using the ACMS-II option. To collect these data, the sample was first zero-field cooled then the data was collected at a fixed temperature for 1 T < $H$ < -1 T. Magnetization versus temperature data were measured in a Quantum Design Magnetic Properties Measurement System (MPMS3) superconducting quantum interference device (SQUID) magnetometer using the vibrating sample magnetometer option. Both field cooled and zero-field cooled data were collected upon warming (stabilizing at each temperature). The collected data were processed by subtracting background signals originating from the substrate using a reference Si substrate. The data were then normalized. For normalization, the film volume was calculated as the product of the area and thickness. The thickness of the films was measured by profilometry, while the area was determined using ImageJ software calibrated against a graph paper grid. The volume determination is certain up to two significant digits.

## DFT calculations

To determine its magnetic ground state, $Co_3ZnN$ was studied across supercells ranging from 5 to 60 atoms, including both cubic, fcc, and tetragonal geometries. For larger supercells (≥ 20 atoms) we performed first-principles Monte Carlo (MC) simulations [33], [37] that used DFT total energies as the acceptance criterion. Simulations were initialized with random moments and run at 100 K [38] with the convergence in 200 MC steps. We also tested AFM and ferrimagnetic (FiM, 1:2 spin-up:spin-down) initial states: both AFM seeds and random-start run relaxed to the same final magnetic state, whereas FiM seeds always relaxed to either FM or AFM, indicating that the FiM configuration is not a locally stable solution. All structures were fully relaxed using spin-polarized GGA for Co-3d in VASP [39], [40], [41] with a 4×4×4 k-point mesh, a 350 eV plane-wave cutoff, and convergence criteria of $10^{-5}$ eV for total energy and 0.02 eV Å$^{-1}$ for forces during DFT Monte Carlo simulations. Final structures from DFT Monte Carlo were subsequently relaxed using more stringent criteria. PAW potentials from VASP 4.6 were employed, including a soft N potential. [42] PDOS calculated in a 5-atom cell using a 12 × 12 × 12 k-point mesh, 380 eV cutoff, and 1 Å integration radius for all Co, Zn, and N atoms.

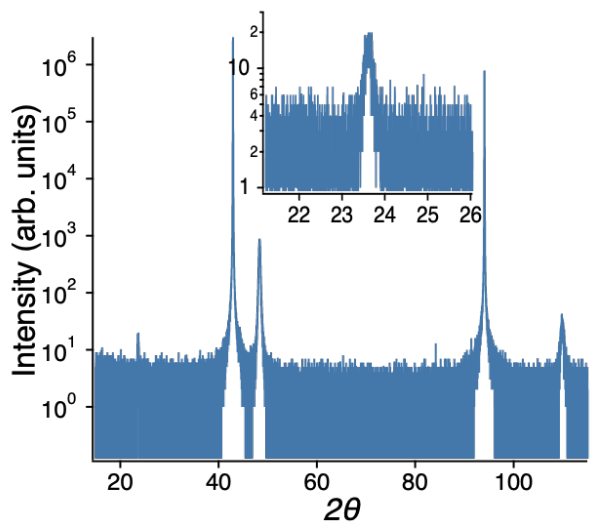

Fig S1: XRD pattern of Co$_3$ZnN on MgO.

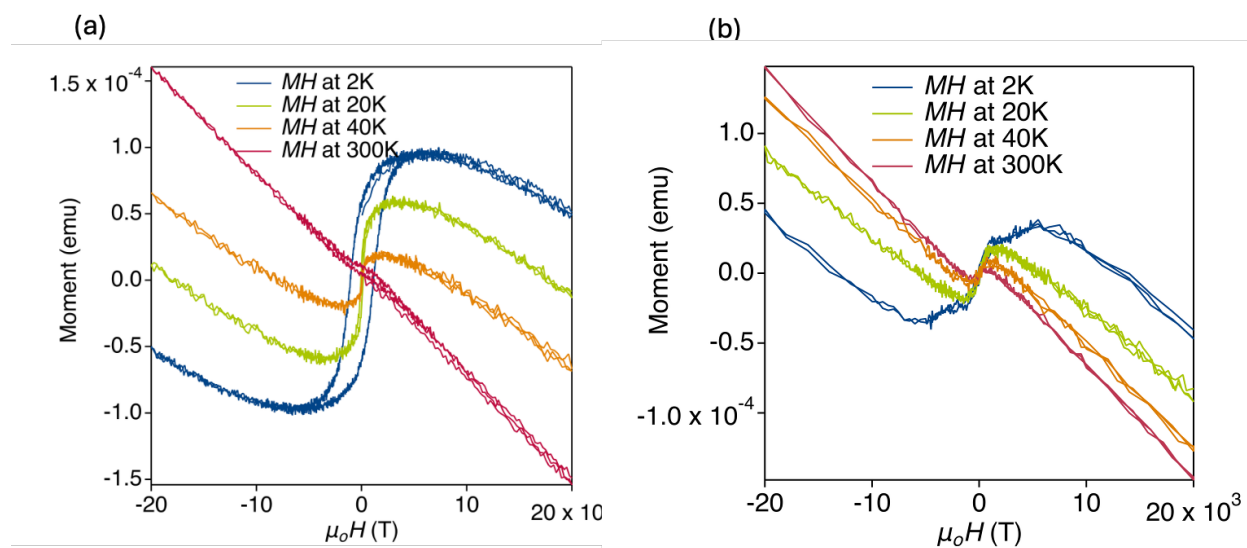

Fig S2: Raw moment data collected as a function of applied magnetic field from 2 K to 300 K for (a) Co$_3$ZnN (with presence of substrate effect) and (b) STO substrate.